# Mid-wave infrared photothermal microscopy for molecular and metabolic imaging in deep tissues and spheroids

Mingsheng Li[1,2,†], Yuhao Yuan[1,2,†], Guangrui Ding[1,2], Hongli Ni[1,2], Biwen Gao[2,3], Dashan Dong[1,2], Qinshu He[1,2], Hongjian He[1,2], Xinyan Teng[2,3], Yuwei Sun[2,4], Dingcheng Sun[2,5], Qing Xia[1,2], Thao Pham[2,5], Ji-Xin Cheng[1,2,3,4,5,*]

1. Department of Electrical and Computer Engineering, Boston University, Boston, MA, USA.
2. Photonics Center, Boston University, Boston, MA, USA.
3. Department of Chemistry, Boston University, Boston, MA, USA.
4. Graduate Program of Molecular Biology, Cell Biology, & Biochemistry, Boston University, Boston, MA, USA.
5. Department of Biomedical Engineering, Boston University, Boston, MA, USA.

† These authors contributed equally to this work.

* Corresponding authors: jxcheng@bu.edu.

## Abstract (149 words)

High-resolution chemical imaging within deep tissues and intact spheroids remains a grand challenge. Here, we introduce mid-wave infrared photothermal (MWIP) microscopy operating in the underexplored 2000–2500 nm spectral window for submicron-resolution molecular and metabolic imaging in intact tumor spheroids and deep tissues. A dark-field photothermal detection scheme significantly suppresses water background and enhances contrast. By accessing strong carbon–hydrogen combination absorptions, a detection limit of 0.12% for dimethyl sulfoxide is achieved, comparable to stimulated Raman scattering microscopy. Depth-resolved imaging of endogenous biomolecules up to 500 μm in excised mouse skin and brain tissues is demonstrated. MWIP further enables depth-resolved tracking of transdermal drug transport via carbon–deuterium overtone absorption. Using deuterium metabolic probes, fatty-acid metabolism is imaged at 200 μm deep within intact tumor spheroids through carbon–deuterium overtone and combination bands. Collectively, MWIP offers a platform for functional imaging of 3D biological systems in their native environments.

# Main

## Introduction

High-resolution chemical imaging of biomolecules and metabolism within deep tissues and intact tumor spheroids is a long-standing goal in biology and medicine[1-3]. Direct visualization of molecular composition and metabolic activity within intact biological systems has shown to be essential for understanding tissue function, disease progression, and therapeutic response[4-6]. Many key biological processes, including drug transport, nutrient uptake, and metabolic reprogramming, are governed by spatial and chemical heterogeneity[7]. However, it still remains a grand challenge to optically penetrate highly scattering biological systems while maintaining molecular specificity and spatial resolution. Near-infrared fluorescence microscopy provides exceptional sensitivity for deep-tissue imaging of molecular and cellular processes[8-10], whereas its broad fluorescence emission spectra provide limited information about molecular structure and compositions. Moreover, fusion with bulky fluorescent tags could introduce systematic artefacts, particularly when tagging small biomolecules.

Providing bond-selective molecular contrast, vibrational spectroscopies, such as infrared (IR)[11,12], confocal Raman[13], and coherent Raman[14-17], have been widely used to study the chemical compositions of biological systems. However, strong water absorption or optical scattering severely restricts their penetration depths to typically <100 μm (**Fig. 1a**).

To interrogate biological machinery in deep tissue, several advanced chemical imaging methods have been explored. Spatially offset Raman spectroscopy can probe millimeter depths by collecting diffusely scattered photons, but its spatial resolution is poor for imaging intracellular functions[18-21]. Probing overtone absorptions, short-wave infrared (SWIR) approaches exploit reduced scattering and attenuated water absorption to extend imaging depth. SWIR diffuse optical imaging employs broadband illumination and analysis of diffusely reflected photons to noninvasively map endogenous chemicals at depths beyond the reach of visible and near-infrared (NIR) methods[22-24]. SWIR photoacoustic microscopy converts optical absorption into ultrasound to enhance penetration depth[25-27]. However, both methods have limited spatial resolution and sensitivity, making them insufficient to resolve subcellular structures in deep tissues. Short-wave infrared photothermal (SWIP) microscopy has enabled micrometer-scale vibrational contrast at millimeter depths using a pump-probe detection scheme[28]. Despite these advances, SWIP imaging relies on relatively weak first carbon–hydrogen (C–H) overtone absorption, limiting detection sensitivity. Moreover, SWIR window does not effectively capture carbon–deuterium (C–D) vibrations, precluding the use of isotopic labeling, which a widely used strategy for probing metabolic activity[29-33].

The mid-wave infrared (MWIR) window, covering the 2000–2500 nm spectral range[34,35], represents an underexplored window for achieving deep-tissue chemical imaging. This window experiences a local minimum of water absorption[36]. Moreover, compared to NIR and SWIR regions, the MWIR window experiences further reduced scattering due to longer wavelengths[37,38]. As shown in **Fig. 1b**, water and tissue phantoms exhibit distinct transmissions across the NIR, SWIR, and MWIR windows. For pure water, MWIR photons have less penetration due to larger water absorption. Yet, owing to the longer wavelengths, MWIR photons show deeper penetration into strongly scattering samples, such as 10% intralipid phantom.

Moreover, MWIR excitation also accesses vibrational transitions highly relevant to molecular content and metabolic activities. This window encompasses strong C–H combination bands as well as C–D overtone and combination transitions (**Fig. 1c**), enabling access to isotope-based metabolic contrast that are inaccessible in NIR and SWIR regimes. Fundamental transitions correspond to excitation from the ground state to the first excited state ($\nu = 0 \rightarrow 1$)[39], whereas the first overtone transitions involve excitation to the second vibrational level ($\nu = 0 \rightarrow 2$)[40-42]. For example, asymmetric $\mathrm{CD_3}$ stretching has a fundamental frequency of ~2210 $cm^{-1}$, and its first overtone lies at ~4420 $cm^{-1}$ (2262 nm) within the MWIR window (**Tables S1, S2**)[43]. The MWIR region also includes combination absorptions of C–H and C–D bonds. Combination absorptions arise when a single photon simultaneously excites multiple fundamental vibrational modes, such as bending and stretching[44]. For instance, 2310-nm excitation simultaneously drives the $\mathrm{CH_2}$ fundamental modes of bending ($\delta\mathrm{CH_2}: 1476$ $cm^{-1}$) and symmetric stretching ($\nu_s\mathrm{CH_2}: 2850$ $cm^{-1}$), which indicates the combination absorption frequency of $4326$ $cm^{-1}$ (2312 nm) ($\delta\mathrm{CH_2} + \nu_s\mathrm{CH_2}: 1476 + 2850 = 4326$ $cm^{-1}$) (**Tables S1, S2**). Unlike C–H overtones which are weak because they are forbidden in the harmonic oscillator approximation, combination bands have stronger absorptions due to coupled vibrational motions that borrow intensity from allowed fundamental modes[40]. Thus, in the MWIR window, C–H combination and C–D overtone/combination absorptions provide strong vibrational signatures of endogenous molecules and deuterium metabolic probes.

Although MWIR vibrational bands of organic molecules have been characterized spectroscopically[43,45], high-resolution MWIR imaging deep within scattering tissues and intact spheroids has not been reported. Here, we introduce mid-wave infrared photothermal (MWIP) microscopy and demonstrate submicron-resolution molecular and metabolic imaging within highly scattering intact tumor spheroids, skin tissue, and brain tissue. MWIP employs a pulsed mid-wave infrared (2000–2500 nm) laser to excite overtone or combination bands and an axially offset continuous-wave probe beam to probe the local thermal lensing effect. Results are detailed below.

# Results

### MWIP microscope with single-pulse digitization and dark-field beam geometry

To extend imaging depth beyond the ballistic photon regime[46], MWIP detects MWIR absorption induced modulations of quasi-ballistic and scattered probe photons while preserving high spatial resolution. To fulfill the MWIP concept, we implemented a pump–probe microscope with single-pulse digitization and a dark-field beam geometry (**Fig. S1**). Pulsed MWIR excitation induces localized nonradiative heating and photothermal lensing, which is detected by an axially offset near-infrared probe in a dark-field configuration. Unlike coherent Raman microscopy[14-17], where the signal is predominantly generated by ballistic photons, MWIP signal arises from MWIR absorption modulation of scattered probe photons and is not limited to ballistic transmission (**Fig. 1d**). As a result, the attainable penetration depth of photothermal imaging is governed by one transport mean free path rather than one single scattering mean free path[37,46], extending the effective imaging depth range to approximately 5 scattering mean free paths[28]. Previous studies have theoretically and experimentally shown that optical resolution can be preserved in the quasi-ballistic photon regime in pump–probe imaging modalities[28,47]. This decoupling of signal generation from ballistic propagation allows the use of a shorter probe wavelength (e.g., 765 nm) to enable submicron spatial resolution without compromising penetration depth.

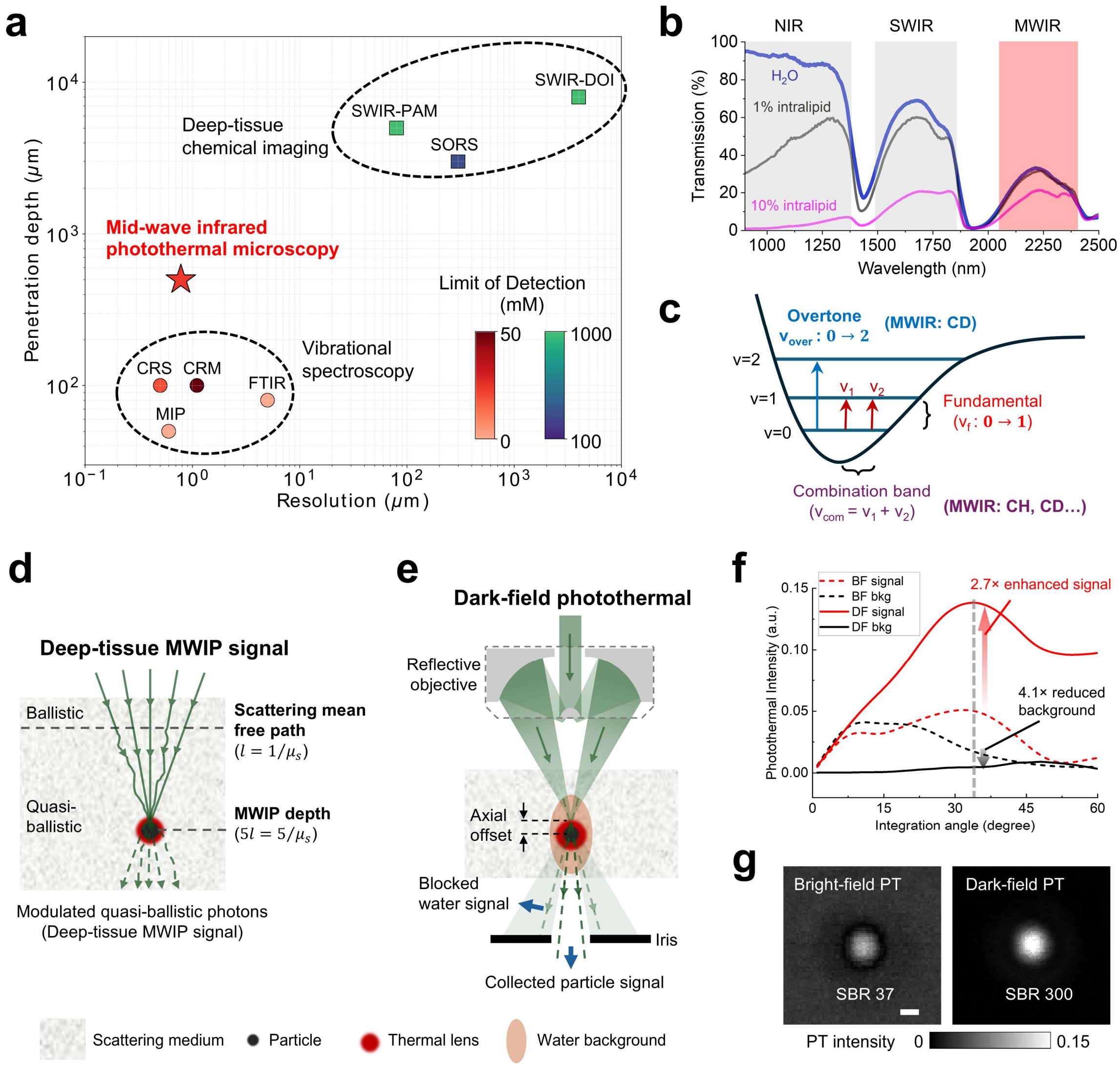

**Figure 1. Mid-wave infrared photothermal (MWIP) microscopy principle and detection scheme. (a)** MWIP microscopy fills the gap between classic vibrational imaging and deep-tissue chemical imaging modalities in terms of spatial resolution, penetration depth, and detection sensitivity. **CRS**, coherent Raman scattering microscopy; **CRM**, confocal Raman microscopy; **MIP**, mid-infrared photothermal microscopy; **FTIR**, Fourier-transform infrared spectroscopy; **SORS**, spatially offset Raman spectroscopy; **SWIR-PAM**, short-wave infrared photoacoustic microscopy; **SWIR-DOI**, short-wave infrared diffuse optical imaging. **(b)** Optical transmission spectra of water and scattering phantom, highlighting near-infrared (NIR), short-wave infrared (SWIR), and mid-wave infrared (MWIR) spectral windows. **(c)** Vibrational energy-level diagram illustrating molecular transitions accessible in the MWIR, including C–D overtone/combination absorptions and C–H combination bands. **(d)** Scheme to show deep-tissue MWIP signal. **(e)** Scheme of dark-field photothermal detection. **(f)** Integrated photothermal signal and water background for bright-field (BF) and dark-field (DF) detection as a function of collection angle. bkg, background. **(g)** Bright-field and dark-field photothermal (PT) microscopic imaging of 1-μm polystyrene bead in water under a scattering phantom using 1725-nm excitation to detect C–H overtone absorption. Details of (**g**) can be found in **Fig. S5**. SBR, signal to background ratio. Scale bar: 1 μm.

Single-pulse digitization enables highly efficient acquisition of low duty cycle photothermal signals[48]. In MWIP, this was achieved by using a fast photodetector and a high-speed digitizer to record the transient response induced by each MWIR pulse at every pixel. Following nonradiative energy deposition, both photothermal and photoacoustic processes are generated[46,49] (**Fig. S2a**). To evaluate the relative detection sensitivity of these two mechanisms within the same excitation, we analyzed both photothermal and photoacoustic contributions from the recorded transient signal. Upon MWIR excitation, a transient photoacoustic component was detected at the onset of the photothermal decay curve (**Fig. S2b**). Notably, the photoacoustic and photothermal responses overlap in both the temporal and frequency domains, such that neither classic time-domain analysis nor frequency-domain filtering alone can reliably resolve the two contributions. To address this limitation, we implemented a wavelet decomposition analysis within the same temporal window to effectively disentangle photoacoustic and photothermal temporal profiles (**Text S1 & Fig. S2c**). The analysis reveals that the photothermal contribution to the refractive index change is approximately 38-fold larger than the photoacoustic contribution. Thus, pump-probe photothermal detection offers a more sensitive mechanism for measuring vibrational absorption than transducer-based photoacoustic detection. Meanwhile, the photoacoustic waves experience much less scattering than optical waves, which allows photoacoustic imaging to reach centimeter-scale penetration depth[50,51].

A major obstacle for chemical imaging in the MWIR window is the water absorption background. Although the MWIR window corresponds to a local minimum in the water absorption spectrum, water remains a dominant background contributor due to its high concentration in biological tissues[23,36]. Even when C–H and C–D absorptions exhibit larger transition dipole moments than water, the volumetric absorption coefficient of water remains substantial because absorption scales with molecular density[46]. As a result, non-specific heating from bulk water generates a significant photothermal background. Furthermore, water exhibits a high specific heat capacity (4.18 J $g^{-1}$ $K^{-1}$), which is significantly larger than other endogenous molecules, such as lipid (1.8–2.3 J $g^{-1}$ $K^{-1}$) and protein (1.3–1.7 J $g^{-1}$ $K^{-1}$)[52]. Thus, for equal absorbed energy, lipids and proteins generate larger local temperature rises than water. However, due to its high molecular concentration, water absorption produces distributed refractive index modulation that forms a substantial photothermal background.

A dark-field photothermal detection scheme is implemented to address the water background issue (**Fig. 1e & Fig. S1**) that was used in the first report of mid-infrared photothermal microscopy[12], but its physical mechanism has not been quantitatively analyzed. Here, we show that this scheme not only suppresses the water background but also enhances the photothermal signal. Dark-field photothermal detection suppresses water background through spatial filtering of the probe beam. Bulk water heating produces refractive index change with a low spatial frequency that only redistributes probe light at small angles. In contrast, localized absorbers such as lipid droplets generate steeper refractive index gradients and broader angular scattering. By rejecting low angle probe photons and collecting higher angle components, dark-field detection suppresses water background while enhancing contrast from localized absorbers.

Finite-element simulations and experiments were performed to compare the dark-field geometry used in this work and the conventional bright-field photothermal detection geometry used in NIR/SWIR overtone photothermal microscopy[28,53] (**Fig. S3**). Angular scattering analysis indicates that dark-field detection enhances the photothermal signal while reducing the water

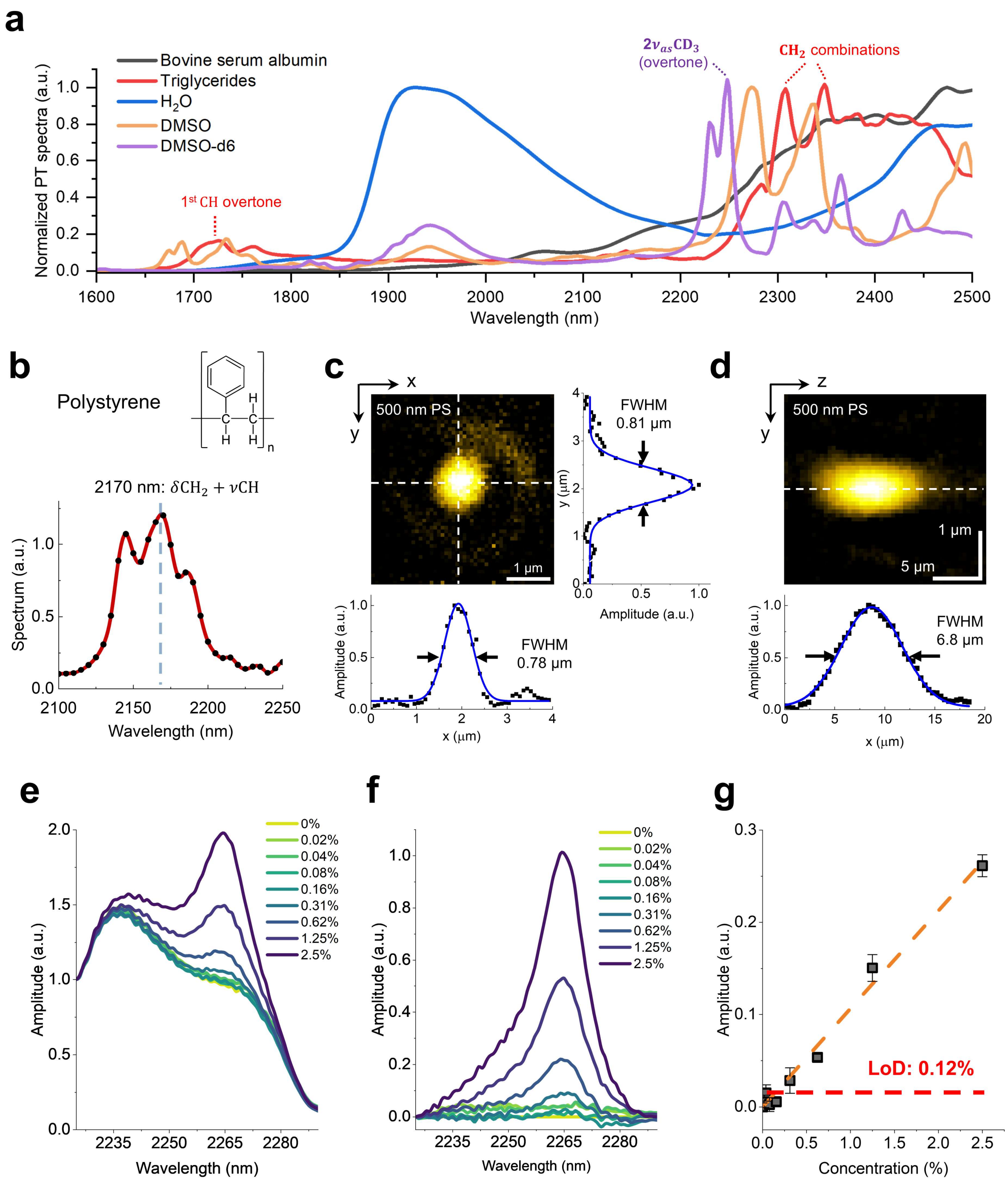

**Figure 2. MWIP spectroscopy and performance of MWIP imaging.**

**(a)** Photothermal spectra of standard biomolecules and solvents showing vibrational signatures. **(b)** Spectrum of a 500-nm PS bead. **(c)** Lateral resolution and **(d)** axial resolution measured by MWIP imaging of a 500-nm PS bead. (**e**) MWIP spectra of DMSO–$D_2O$ mixtures at varying concentrations. **(f)** MWIP spectra with background subtracted. **(g)** Calibration curve shows a detection limit of 0.12%. DMSO, dimethyl sulfoxide. PS, polystyrene. LoD, limit of detection.

background (**Fig. S4**). Integration of the far-field intensity distribution reveals a 2.7-fold signal enhancement and a 4.1-fold background suppression for dark-field detection compared with conventional bright-field method (**Fig. 1f**). To experimentally validate this improvement, bright-field and dark-field photothermal images of a 1-μm polystyrene bead in water beneath a scattering phantom were acquired by probing C–H overtone absorption (**Fig. 1g & S5**). Dark-field detection yields approximately an order-of-magnitude improvement in signal-to-background ratio compared with conventional bright-field approach.

Together, these results demonstrate the feasibility of using MWIP microscopy for deep-tissue chemical imaging with reduced water background and improved contrast.

**MWIP spectroscopy and performance of MWIP imaging**

To unveil the molecular signatures in MWIR spectral window, MWIP spectra of standard samples were measured and compared with the short-wave infrared photothermal spectra (**Fig. 2a**). For pure water, a local minimum was observed in the MWIP window from 2100 to 2400 nm. The first overtone absorption of asymmetric $\mathrm{CD_3}$ stretching mode in dimethyl sulfoxide-d6 (DMSO-$d_6$) was observed at 2262 nm, while combination band absorptions of $\mathrm{CH_2}$ groups in triglycerides were detected at 2310 nm and 2350 nm. A broad band was observed from albumin proteins. Additional peak assignments are provided in **Table S1**. The intensity of the $\mathrm{CH_2}$ combination band is approximately 7-fold higher than that of the first C–H overtone at 1725 nm due to a larger absorption cross section[54]. Together, these vibrational signatures build the foundation for MWIP imaging of endogenous biomolecules and isotope-labeled metabolites.

The use of a NIR probe wavelength (765 nm) in our system enables submicron spatial resolution and optical sectioning. To quantify the spatial resolution, MWIP spectroscopic imaging was performed on a 500-nm polystyrene (PS) particle (**Fig. 2b**). The MWIP spectrum exhibits a peak at 2170 nm corresponding to the C–H combination band ($\delta\mathrm{CH_2} + \nu\mathrm{CH}$) of PS (**Table S1**). The lateral resolution, determined from the full width at half maximum (FWHM) of the intensity line profile, was measured as 0.78 μm and 0.81 μm along orthogonal directions (**Fig. 2c**). Depth-resolved imaging of the same particle yielded an axial resolution of 6.8 μm, defined by the FWHM of the axial intensity profile (**Fig. 2d**). Detection sensitivity was evaluated using dimethyl sulfoxide (DMSO) solutions in deuterated water as a testbed (**Fig. 2e**). After subtraction of the solvent background, the residual spectra show concentration dependent absorption features (**Fig. 2f**). The MWIP signal amplitude obtained by spectral integration exhibits a linear relationship with DMSO concentration (**Fig. 2g**). The limit of detection (LoD), defined as a signal equal to three times the standard deviation of the noise, was determined to be 0.12% (**Text S2**). This value is comparable to reported detection limits for the well-established stimulated Raman scattering microscopy[55,56] for which the LoD of DMSO is ~0.1%. Together, these results show that MWIP allows submicron-resolution chemical imaging with a high sensitivity.

**High-resolution MWIP imaging of endogenous molecules in excised mouse skin and brain**

Visualizing endogenous molecular composition in deep tissues is important for understanding chemical heterogeneity under physiological conditions, whereas optical scattering and water absorption limit the penetration depth of classic vibrational spectroscopic imaging methods[11,57-60]. We therefore evaluated MWIP for imaging natural biomolecules in highly scattering tissues and benchmarked its performance against SWIP microscopy[28]. **Fig. 3a-b** show SWIP (1725 nm) and

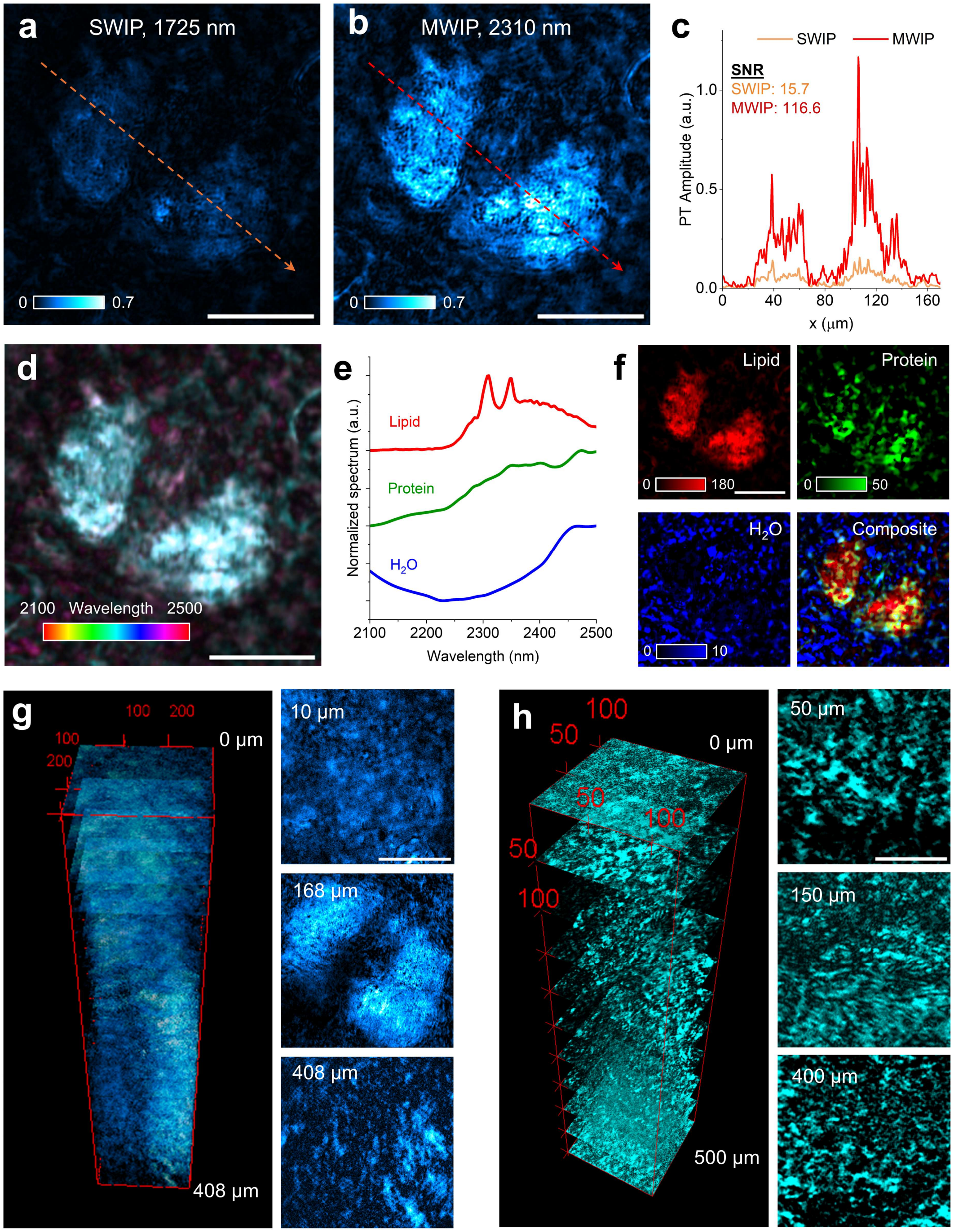

**Figure 3. High-resolution MWIP imaging of endogenous molecules in excised mouse skin and brain.**

**(a)–(b)** Benchmarking of short-wave infrared photothermal (SWIP) and mid-wave infrared photothermal (MWIP) microscopes by imaging a same sebaceous gland at a depth of 130 μm under mouse skin. **(c)** Line profiles show MWIP presents an enhanced 8-fold SNR than SWIP. **(d)** Pseudo-color hyperspectral image of sebaceous glands and surrounding tissue. **(e)** Reference spectra of lipid, protein, and water. **(f)** Spectral unmixing maps of lipid, protein, and water, and composite image. **(g)** 3D MWIP reconstruction of mouse skin with depth-resolved slices using excitation wavelength 2310 nm to show lipid contrast. **(h)** 3D reconstruction and depth sections of a 500-μm thick mouse brain slice using excitation wavelength 2310 nm to show lipid contrast. Scale bar: (a-g) = 50 μm, h = 25 μm.

MWIP (2310 nm) images of lipids in a sebaceous gland within excised mouse ear skin at a depth of 130 μm, acquired over the same field of view. SWIP contrast arises from the first C–H overtone absorption[36], whereas MWIP probes the C–H combination band which has a larger absorption cross section. Line profile analysis across the gland structure (**Fig. 3c**) indicates an approximately 8-fold higher signal-to-noise ratio (**Text S3**) between MWIP and SWIP under the same illumination power conditions, consistent with the spectra shown in **Fig. 2a**.

Further, hyperspectral MWIP imaging combined with computational spectral unmixing was performed to resolve multiple molecular components within the tissue. **Fig. 3d** shows a pseudo-color hyperspectral image of sebaceous glands and surrounding tissue, acquired across the MWIR spectral range. Pixel-wise spectral unmixing was performed using a least absolute shrinkage and selection operator (LASSO) algorithm[61] with reference spectra from pure chemical components (**Fig. 3e**). The resulting chemical maps (**Fig. 3f**) distinguish lipid, protein, and water signals. The sebaceous gland region exhibits dominant lipid contrast, consistent with its known lipid-rich composition[62,63].

Next, depth-resolved MWIP imaging was performed to evaluate penetration performance in scattering tissues. Using 2310 nm excitation, lipid contrast was visualized at multiple depths in excised mouse ear skin (**Fig. 3g**). At a shallow depth of 10 μm, the stratum corneum exhibited a layered morphology corresponding to corneocytes and intercellular lipids[64]. At the depth of 168 μm within the epidermal layer, sebaceous glands remained clearly resolved with preserved spatial contrast. Lipid signals were still detectable at depths of 408 μm, with a measured signal-to-noise ratio of 36. In brain tissue, depth-resolved MWIP imaging of a 500-μm thick brain slice revealed submicron lipid features at depths up to 400 μm using 2310 nm excitation (**Fig. 3h & Fig. S6**). Structural patterns consistent with myelinated regions were observed at a depth of 150 μm. These results indicate that MWIP maintains molecular contrast and sub-micron spatial resolution at sub-millimeter depths in highly scattering tissues, significantly extending beyond the penetration capability of classic vibrational spectroscopic imaging[11,57-60], which is typically <100 μm.

Collectively, these results show that MWIP provides higher sensitivity than recently reported SWIP[28] and enables high-resolution chemical imaging at much greater depths than linear or nonlinear vibrational spectroscopy methods shown in **Fig. 1a**.

**MWIP investigation of transdermal drug penetration with high spectral fidelity**

High-fidelity spectroscopic imaging in deep tissue is crucial for evaluating transdermal drug delivery and pharmacokinetics, where accurate molecular specificity and depth-resolved measurements are required to link drug distribution with therapeutic efficacy[65,66]. Here, we assessed the spectral fidelity of MWIP in scattering medium and applied it to depth-resolved imaging of transdermal drug penetration in excised mouse ear skin.

The spectral fidelity of MWIP was evaluated using a scattering phantom. A 500-μm thick scattering phantom composed of 1% intralipid in agar was used to mimic the scattering tissue[67,68] (**Fig. 4a**). An 80-μm thick DMSO-$d_6$ layer was placed beneath the phantom, then its MWIP spectrum was measured and compared with the reference spectrum acquired without scattering. As shown in **Fig. 4b**, the spectral features of DMSO-$d_6$ were preserved after propagation through the scattering layer, indicating that MWIP maintains high-fidelity spectroscopic contrast under highly scattering conditions. Photothermal detection is the foundation for such spectral fidelity, where MWIR absorption is measured by a single-wavelength probe beam.

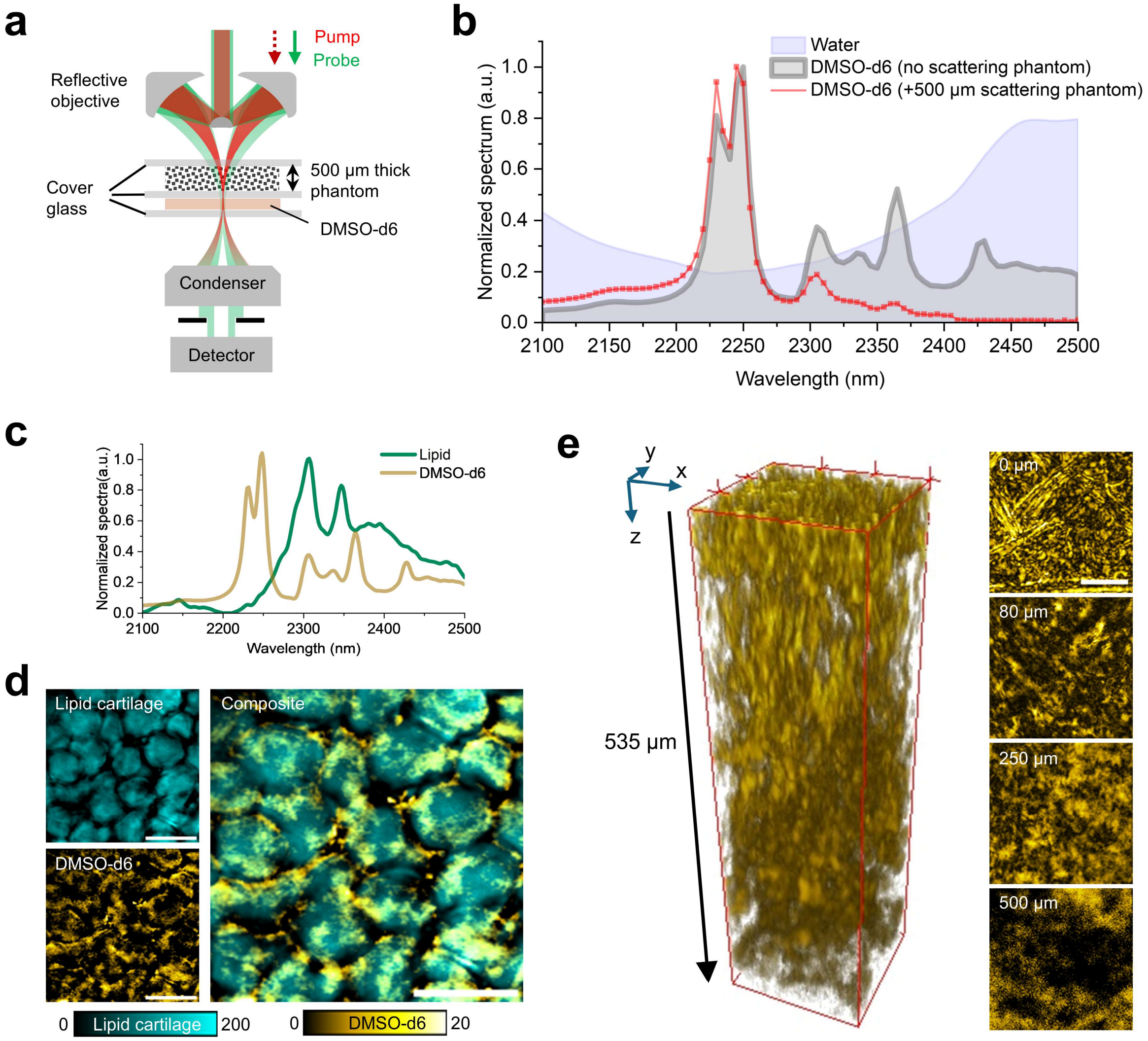

**Figure 4. MWIP investigation of transdermal drug penetration with high spectral fidelity.** **(a)** Scheme of spectroscopic measurement of DMSO-$d_6$ under 500-μm thick 1% intralipid scattering phantom. **(b)** MWIP spectra of DMSO-$d_6$ with and without scattering phantom. **(c)** Reference spectra of lipid and DMSO-$d_6$ (model drug compound) for LASSO spectral unmixing. **(d)** Spectral unmixing maps and composite image showing lipid cartilage and drug distributions. **(e)** 3D reconstruction of DMSO-$d_6$ penetration through a 535-μm skin volume, and depth-resolved images at selected depths (0–500 μm). Scale bars: d = 30 μm; f = 50 μm. DMSO-$d_6$, dimethyl sulfoxide-d6.

We then applied MWIP to visualize transdermal drug penetration into a mouse skin. DMSO-$d_6$ was used as a model compound because DMSO is commonly used as a drug vehicle to enhance transdermal penetration efficacy[69]. Hyperspectral MWIP imaging of excised mouse ear skin tissue revealed composite spectral features from lipid cartilage and drug content at a depth of 180 μm beneath the surface (**Fig. S7**). Pixel-wise spectral unmixing using a LASSO algorithm[61] separated lipid cartilage and DMSO-$d_6$ signals based on reference spectra (**Fig. 4c**). The resulting maps in **Fig. 4d** show drug compound localized within intercellular spaces of lipid cartilage structures. Depth-resolved MWIP imaging of drug distribution in skin is shown in **Fig. 4e**. Drug contrast remained detectable to a maximum depth of 535 μm, which is beyond the reach of classic vibrational spectroscopy, such as IR[11,12] and coherent Raman[55,70].

Collectively, these results establish MWIP as a platform for high-fidelity, high-resolution spectroscopic imaging in 500 μm deep tissues, enabling visualization of transdermal drug transport through skin layers.

**Imaging fatty acid metabolism in OVCAR5 tumor spheroids via probing C–D**

Tumor spheroids recapitulate the three-dimensional architecture and metabolic heterogeneity of solid tumors, where nutrient metabolism is influenced by diffusion barriers and local microenvironments[71-73]. However, classic vibrational spectroscopies are limited to shallow depths (~50 μm) in dense spheroids due to scattering or water attenuation[74,75], and short-wave infrared methods do not effectively access the vibrational features of fatty acid metabolism[28]. We therefore evaluated MWIP for high-resolution depth-resolved metabolic imaging within intact OVCAR5 tumor spheroids by probing C–D overtone and combination absorptions (**Table S1**).

To trace fatty acid metabolism, palmitic acid-$d_{31}$ (PA-$d_{31}$) was used as a deuterium metabolic probe for isotopic labelling (**Fig. 5a**). MWIP spectral imaging was first performed in two-dimensional OVCAR5 cell cultures as a testbed to validate deuterium metabolic tracing (**Fig. 5b**). Cells were incubated with either unlabeled palmitic acid (PA) or PA-$d_{31}$ for 24 hours. Summed hyperspectral images over the 2200–2500 nm range show cell morphology, while pixel-wise LASSO spectral unmixing[61] separates CH and CD signals. The CD map shows newly synthesized lipids in PA-$d_{31}$ treated cells, whereas control cells exhibit only CH lipid contrast showing the total lipid amount. Automated cell segmentation using Cellpose[76] (**Fig. S8**) enabled quantitative analysis of individual cells. In **Fig. 5c**, the CD/CH intensity ratio is used to quantify fatty acid uptake, demonstrating metabolic contrast based on deuterium incorporation. These results validate the capability of MWIP microscopy to image fatty acid metabolism by spectral tracing of deuterium through isotope labelling.

High-resolution MWIP imaging of fatty acid metabolism deep within intact OVCAR5 tumor spheroids is shown in **Fig. 5d**. Control and PA-$d_{31}$ treated spheroids were incubated for 24 hours to allow efficient nutrient uptake. Summed hyperspectral images show overall spheroid morphology, followed by LASSO spectral unmixing[61] to separate the CH and CD signals. In control spheroids, the CH map represents total lipid content, while CD contrast is not observed. In PA-$d_{31}$ treated spheroids, CD maps reveals newly synthesized lipids. At a shallow depth of 60 μm, MWIP efficiently visualize the lipid (CH) and CD maps due to minimal attenuation of photons. The CD image exhibits relatively uniform contrast across all regions of the spheroids, indicating efficient uptake of fatty acids into the superficial regions of the spheroids. At a depth of 140 μm, MWIP maintains morphological and chemical contrast as shown in CH image. However, CD signal intensity decreases within spheroid compared with the periphery, indicating reduced fatty acid uptake toward the spheroid interior. At a depth of 200 μm, signals remain detectable primarily in peripheral regions, while CD contrast further decreases toward the spheroid center due to increased optical attenuation and reduction of fatty acid uptake. Thus, our results determine the penetration depth limit of MWIP metabolic imaging into intact tumor spheroids as 200 μm. The CD and CH maps at depths of 140 and 200 μm show a spatial gradient of fatty acid uptake from the periphery toward the center, consistent with diffusion-limited nutrient transport in dense spheroids[77].

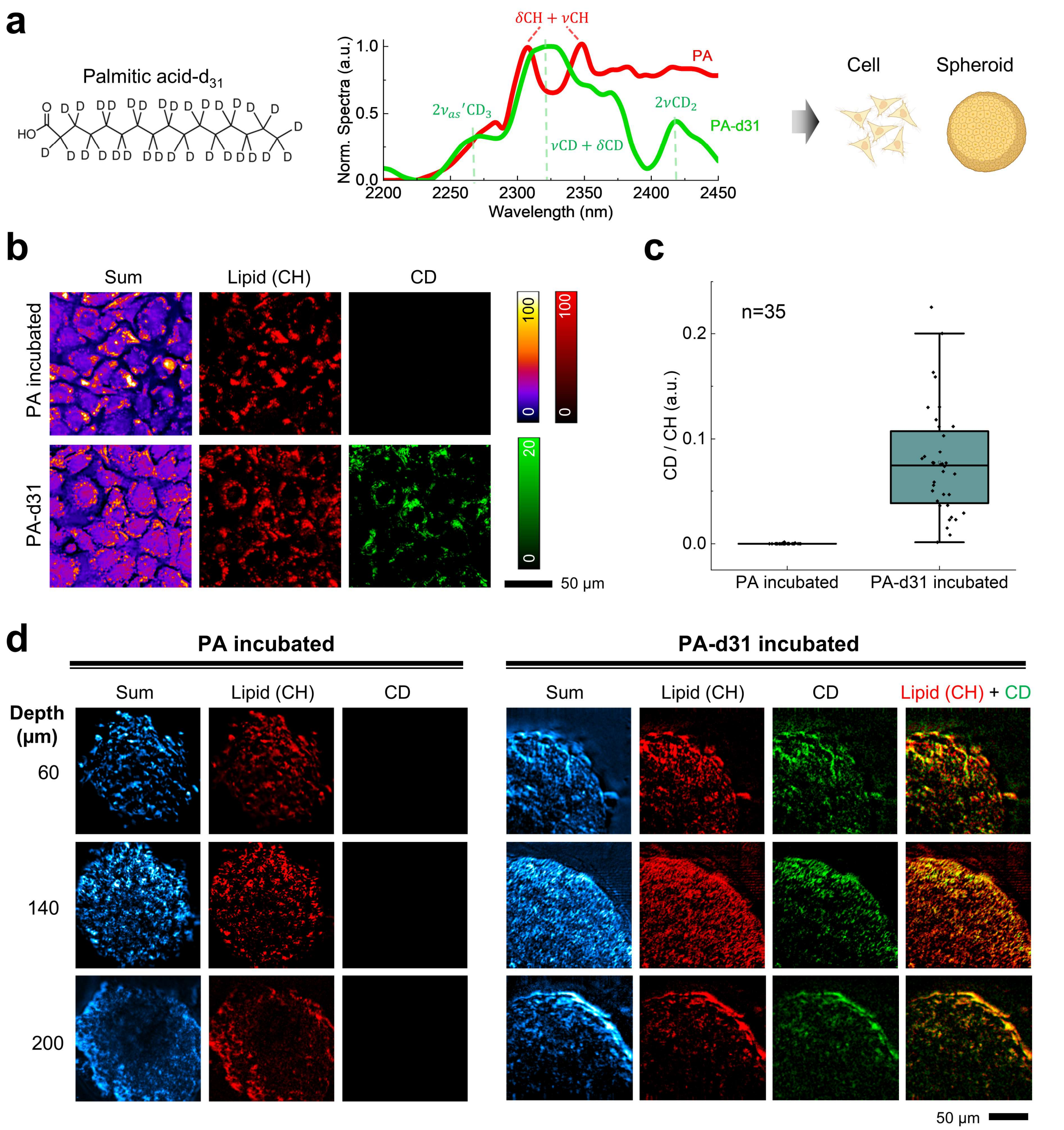

**Figure 5. Imaging fatty acid metabolism in OVCAR5 tumor spheroids via probing C–D**.
**(a)** Schematic of PA-$d_{31}$ labeling and spheroid formation, and representative spectra showing CH and CD absorption signatures. **(b)** Sum of hyperspectral, lipid (CH), and CD images of PA and PA-$d_{31}$ incubated cells. **(c)** CD/CH ratio quantification (n = 35) after cell segmentation. **(e)** Depth-resolved imaging of PA and PA-$d_{31}$ incubated OVCAR5 tumor spheroids. PA, palmitic acid. PA-$d_{31}$, palmitic acid-d31.

Collectively, these results demonstrate that MWIP allows depth-resolved metabolic imaging of stable isotope probes (e.g., PA-d31) in intact tumor spheroids, with penetration capability beyond the reach of classic vibration spectroscopy, which is limited to ~50 μm[74,75].

**Discussion**

Mid-wave infrared photothermal (MWIP) microscopy establishes a platform for high-sensitivity molecular and metabolic imaging with extended penetration depth and submicron spatial resolution. Classic vibrational spectroscopies provide rich molecular specificity but are restricted to shallow depths[11,58-60,78,79] , whereas deep-tissue optical approaches generally sacrifice spatial resolution, sensitivity, or chemical specificity[18-20,22-28]. MWIP fills this capability gap by leveraging MWIR vibrational absorption with photothermal detection, enabling chemical imaging in regimes not accessible to previous approaches.

The MWIR window (2000–2500 nm) is an underexplored regime for chemical imaging, providing rich molecular vibrational absorption features that are inaccessible in NIR and SWIR windows[43,45]. C–H combination bands provide larger absorption cross sections than C–H overtones, enabling an improved sensitivity. The C–D overtone and combination transitions enable metabolic imaging through deuterium labeling. The overtone and combinational transitions of nitrile and azide also reside in the window[80-82], which heralds even broader applications of MWIP microscopy. On the instrumentation side, dark-field photothermal detection scheme significantly suppresses water background and enhances the signal. Owing to the large cross sections and the advanced detection scheme, MWIP enables highly sensitive molecular imaging in scattering biological media while maintaining a high spatial resolution.

The penetration depth of MWIP microscopy is limited by scattering of probe photons and the water attenuations of pump photons. Unlike coherent Raman microscopy in which the penetration depth is determined by one scattering mean free path ($l = 1/\mu_s$, $\mu_s$ is scattering coefficient), MWIP can operate in the quasi-ballistic photon regime, and its penetration is governed by one transport mean free path ( $l' = 1/\mu'_s = \frac{1}{\mu_s(1-g)} \approx 10\ l$ , where $\mu'_s$ is the reduced scattering coefficient, and $g$ is the anisotropy factor ~0.9)[37]. In practice, due to the photon collection and laser noise limitations, deep-tissue photothermal detection can operate at depth of up to 5 scattering mean free paths. For instance, the scattering coefficients ($\mu_s$) of 765 nm and 1310 nm probe laser used in MWIP and previously reported SWIP[28] systems are approximately 10 $\mathrm{mm}^{-1}$ and 5 $\mathrm{mm}^{-1}$ in brain tissue. These values correspond to scattering mean free paths ($1/\mu_s$) of 100 μm and 200 μm. Assuming a practical operating depth of 5 mean free paths, the expected penetration limits are 500 μm for MWIP and 1 mm for SWIP. These theoretical considerations provide a quantitative framework to explain the experimentally observed penetration depths of both MWIP and SWIP systems, which agree well with prior SWIP studies[28]. Furthermore, this analysis suggests that the penetration depth of MWIP can be further extended by employing longer probe wavelengths (e.g., 800–1000 nm), which would reduce scattering while still maintaining high spatial resolution.

Overall, MWIP microscopy establishes a framework for submicron-resolution chemical imaging of molecular compositions and metabolic activity in highly scattering intact tumor spheroids and deep tissues. Being able to access C–H combination bands as well as overtone and combination transitions of C–D, nitrile, and azide groups, MWIP expands the vibrational imaging toolbox beyond vibrational spectroscopies and conventional NIR/SWIR approaches. This capability opens new opportunities for probing molecular composition, transdermal drug transport, and metabolic dynamics in complex biological systems with both high chemical specificity and extended imaging depth.

## Methods

### 1. A mid-wave infrared photothermal microscope

**Fig. S1a** shows the schematic of experimental set up of MWIP microscope. A tunable pump laser (Ekspla OPO, 600–2600 nm) is combined with a continuous-wave probe laser (765 nm, TLB6712-D, Spectra Physics) using a dichroic mirror. A 4f system with 50-mm and 100-mm $CaF_2$ lenses (LA5763 and LA5817, Thorlabs) provides axial offset control at the focal plane. The collimated beams are focused onto the sample by a reflective objective (LMM40X-P01, Thorlabs). Transmitted probe light is collected by a condenser, spatially filtered by an iris, and spectrally filtered by a short-pass filter. A photodiode converts the detected photons into electrical signals, which are amplified by a low-noise voltage amplifier (100 MHz bandwidth, SA230F5, NF Corporation) and digitized by a high-speed acquisition card at 180 MSa/s (ATS9462, Alazar Tech). As shown in **Fig. S1b**, for each pump pulse, a time-resolved transient signal was recorded. Photothermal contrast was defined as the difference between the probe intensity before pumping, corresponding to the thermal equilibrium state, and the peak probe intensity during the thermal transient, corresponding to the maximum temperature rise. Images were generated by raster scanning the sample using a motorized stage (Nano-Bio 2200, Mad City Labs).

### 2. Finite-element simulations by COMSOL

Finite-element simulations were performed in COMSOL Multiphysics to model photothermal detection. A 500-nm polystyrene particle embedded in water was used as the model absorber, with identical absorbed photon energy assigned to both the particle and surrounding medium. For background analysis, the particle was removed and only the photothermal response of water was considered. “Hot” and “cold” states correspond to the heating and cooling phases during photothermal modulation. Far-field angular scattering distributions were computed for both bright-field and dark-field detection geometries. The photothermal signal was quantified by integrating the differential radiative intensity (hot − cold) over the collection angles.

### 3. Spectral unmixing via pixel-wise LASSO

Spectral unmixing via pixel-wise least absolute shrinkage and selection operator (LASSO) was applied to hyperspectral MWIP image datasets to quantitatively generate chemical maps[53,61,83]. For each pixel, the measured MWIP spectrum was modeled as a linear combination of reference spectra corresponding to major biochemical components. Reference spectra were obtained from pure samples or regions of interest dominated by individual components and normalized prior to unmixing. Pixel-wise LASSO regression was then performed by minimizing a least-squares error term with an L1-norm sparsity constraint, enforcing the assumption that only a limited number of chemical species contribute significantly at each spatial location. The regularization parameter was empirically optimized to suppress spectral crosstalk while preserving spatial fidelity. The resulting concentration coefficients were used to reconstruct quantitative chemical maps for individual molecular components, which were subsequently visualized and analyzed.

## 4. Brain slices preparations

The brain tissue used in this study was obtained from a 2-month-old BALB/c mouse. Following sacrifice, the mouse was perfused with phosphate-buffered saline (PBS, 1×; Thermo Fisher Scientific), followed by 10% formalin. The brain was subsequently extracted and post-fixed in 10% formalin for 48 h. Coronal brain slices (500 µm thick) were prepared using an oscillating tissue slicer (OTS-4500, Electron Microscopy Sciences) and washed three times with PBS prior to imaging. For MWIP imaging, the brain tissue was immersed in PBS and mounted between two cover glasses with a spacer.

## 5. Cell and tumor spheroid culture

OVCAR5 human ovarian cancer cells were cultured in Dulbecco's Modified Eagle Medium (DMEM, Thermo Fisher Scientific) supplemented with 10% (v/v) fetal bovine serum (FBS, Gibco) and 1% (v/v) penicillin–streptomycin. Cells were maintained at 37 °C in a humidified incubator with 5% $CO_2$ and routinely passaged using trypsin–EDTA before reaching 80% confluency.

For two-dimensional (2D) cell culture experiments, OVCAR5 cells were seeded onto glass coverslips at an appropriate density and allowed to adhere overnight prior to metabolic labeling. For fatty acid uptake studies, cells were incubated with either unlabeled palmitic acid (PA) or deuterium-labeled palmitic acid-d31 (PA-$d_{31}$) at a concentration of 50 µM for 24 h in complete culture medium. After incubation, cells were washed three times with phosphate-buffered saline (PBS), fixed with 10% formalin, and washed an additional three times with PBS prior to imaging.

Three-dimensional (3D) tumor spheroids were generated using a low-adhesion culture approach. Briefly, OVCAR5 cells were seeded into ultra-low-attachment round-bottom 96-well plates (Corning) at a defined cell number per well to promote spheroid formation via self-aggregation. Spheroids were allowed to form and mature for 24 h until compact and dense spheroids were obtained. Mature spheroids were then incubated with PA or PA-$d_{31}$ at a concentration of 50 µM for 24 h to allow fatty acid uptake under diffusion-limited conditions. Following incubation, spheroids were gently washed with PBS to remove excess extracellular fatty acids, fixed with 10% formalin, washed three times with PBS before imaging. All spheroid handling steps were performed with minimal mechanical perturbation to preserve structural integrity.

## 6. Automated single-cell segmentation by Cellpose

Automated single-cell segmentation was performed using Cellpose, a generalist deep-learning–based segmentation algorithm[76]. Hyperspectral images were first preprocessed by summing signal intensities across selected spectral channels to generate a high-contrast input image for segmentation. Cell boundaries were then identified using the pretrained Cellpose model without additional retraining. The algorithm predicts pixel-wise vector flow fields and cell probability maps, which were used to generate instance segmentation masks for individual cells. Segmented cell masks were visually inspected to ensure accurate boundary delineation and subsequently used for downstream quantitative analysis.

**Acknowledgement**
This work was supported by R35GM136223 to JXC. The authors thank Danchen Jia for helpful discussion on COMSOL simulations.

**Contributions**
J.-X.C., Y.Y., and M.L. conceived the study and designed the experiments. Y.Y., M.L. and H.N. developed the setup. M.L. and Y.Y. performed the experiments and conducted data analysis. G.D. contributed to spectral unmixing. B.G. assisted with instrumentation of the probe path. D.D. and Q.H. performed the COMSOL simulations. X.T., H.H., Y.S., and D.S. contributed to sample preparation. Q.X. assisted with data analysis, and T.P. contributed to conducting experiments. J.-X.C. supervised the project and provided overall guidance. M.L., Y.Y., and J.-X.C. drafted the manuscript. All authors contributed to the manuscript through feedback and revisions.

# Supplementary Information

## Mid-wave infrared photothermal microscopy for molecular and metabolic imaging in deep tissues and spheroids

Mingsheng Li[1,2,†], Yuhao Yuan[1,2,†], Guangrui Ding[1,2], Hongli Ni[1,2], Biwen Gao[2,3], Dashan Dong[1,2], Qinshu He[1,2], Hongjian He[1,2], Xinyan Teng[2,3], Yuwei Sun[2,4], Dingcheng Sun[2,5], Qing Xia[1,2], Thao Pham[2,5], Ji-Xin Cheng[1,2,3,4,5,*]

1. Department of Electrical and Computer Engineering, Boston University, Boston, MA, USA.
2. Photonics Center, Boston University, Boston, MA, USA.
3. Department of Chemistry, Boston University, Boston, MA, USA.
4. Graduate Program of Molecular Biology, Cell Biology, & Biochemistry, Boston University, Boston, MA, USA.
5. Department of Biomedical Engineering, Boston University, Boston, MA, USA.

† These authors contributed equally to this work.

* Corresponding authors: jxcheng@bu.edu.

**This PDF file includes**

**Supplementary Texts (Text S1 to S3)**

**Supplementary Tables (Table S1 to S2)**

**Supplementary Figures (Figure S1 to S8)**

# Supplementary Texts

## 1. Wavelet decomposition to differentiate photothermal and photoacoustic signals

Pulsed mid-wave infrared excitation generates a composite transient signal consisting of a fast photoacoustic (PA) component on the nanosecond timescale and a slower photothermal (PT) component on the microsecond timescale (**Fig. S2**). The measured signal can be expressed as

$$s(t) = s_{PA}(t) + s_{PT}(t)$$

To separate these components, a maximal overlap discrete wavelet transform (MODWT) with a fourth-order Fejér–Korovkin wavelet was applied. The signal was decomposed into wavelet detail coefficients $W_j(t)$and a scaling coefficient $V_J(t)$,

$$s(t) = \sum_{j=1}^{J} W_j(t) + V_J(t)$$

High-frequency PA transients are captured by lower-level coefficients, while the slower PT response is contained in higher-level coefficients and $V_J(t)$ . Accordingly, the reconstructed PA and PT signals were defined as

$$s_{\mathrm{PA}}(t) = \sum_{j=1}^{j_c} W_j\,(t)$$

$$s_{\mathrm{PT}}(t) = \sum_{j=j_c+1}^{J} W_j\,(t) + V_J(t)$$

where $j_c$ denotes the cutoff level separating fast and slow dynamics.

The PA amplitude was quantified by integrating the Hilbert envelope over the PA duration (1). And the PT amplitude was obtained by integrating the reconstructed PT component over the same window.

## 2. Determine limit of detection

The limit of detection (LoD) of the MWIP system was quantified based on the signal-to-noise characteristics of the photothermal signal. For a given vibrational band, the photothermal signal $S$ was obtained by integrating the time-resolved photothermal intensity over the spectral window of interest,

$$S = \int_{\lambda_1}^{\lambda_2} I_{\mathrm{PT}}\,(\lambda)\, d\lambda.$$

The noise level $\sigma$ was defined as the standard deviation of the integrated signal measured from solvent-only or background regions. A linear calibration between signal amplitude and analyte concentration $C$ was established as

$$S = kC + S_0,$$

where $k$ is the calibration slope and $S_0$ is the background offset. The LoD was defined using the standard criterion

$$\mathrm{LoD} = \frac{3\sigma}{k},$$

corresponding to a signal exceeding the noise floor with high confidence. Using this method, the LoD for DMSO in deuterated water was determined to be 0.12%.

### 3. Determine signal to noise ratio

The signal-to-noise ratio (SNR) of MWIP images was evaluated to quantify imaging sensitivity at different depths. The signal $S$ was defined as the mean photothermal intensity within a region of interest (ROI) containing the target structure,

$$S = \frac{1}{M}\sum_{i=1}^{M} I_i\,,$$

where $I_i$ is the photothermal signal of the $i$-th pixel in the ROI. The noise $\sigma$ was calculated as the standard deviation of the signal measured from a nearby background region,

$$\sigma = \sqrt{\frac{1}{N-1}\sum_{j=1}^{N}\left(I_j - \bar{I}_{\mathrm{bg}}\right)^2}\,.$$

The SNR was then calculated as

$$\mathrm{SNR} = \frac{S}{\sigma}.$$

# Supplementary Tables

## Supplementary Table 1. Mid-wave infrared spectral window: combinational bands and overtone vibrational modes and peak assignment.

| Vibration modes | Description | Wavenumber ($cm^{-1}$) | Wavelength (nm) | Reference |
|---|---|---|---|---|
| $\nu C=C + \nu CH$ | Combinational bands | 4648 | 2150 | (*2*) |
| $\delta CH_2 + \nu CH$<br>$\delta CH_2 + \nu_{as} CH_3$ | Combinational bands | 4606 | 2171 | (*2*) |
| $\delta CH_2 + \nu_{as} CH_2$<br>$\delta CH_2 + \nu_s CH_2$<br>$\delta CH_2 + \nu_s CH_3$ | Combinational bands | 4329 | 2310 | (*2*) |
| $(\delta CH_2, \delta OH) + \nu_s CH_2$ | Combinational bands | 4245 | 2356 | (*2*) |
| $2\nu_{as} CD_3$ | 1st overtone of asymmetric $CD_3$ stretching | 4420 | 2262 | (*3*) |
| $2\nu_{as}' CD_3$ | 1st overtone of the degenerate asymmetric $CD_3$ stretching | 4403 | 2271 | (*3*) |
| $\nu CD + \delta CD$ | C–D stretch and bend combination band | ~4237-4292 | ~2330-2360 | (*4*) |
| $\nu_s CD_3 + \nu_{as} CD_3$ | Combinational bands | 4340 | 2304 | (*3*) |
| $\nu_{as}' CD_3 + \nu_s CD_3$ | Combinational bands | 4275 | 2339 | (*3*) |

**Supplementary Table 2. Infrared spectral window: fundamental transition modes and peak assignment.**

| Vibration modes | Description | Wavenumber ($cm^{-1}$) | Reference |
|---|---|---|---|
| $\nu_s CH_2$ | Symmetric $CH_2$ stretching | 2850 | (*5, 6*) |
| $\nu_{as} CH_2$ | Asymmetric $CH_2$ stretching | 2925 | (*7*) |
| $\delta CH_2$ | $CH_2$ bending | 1476 | (*8*) |
| $\nu_s CH_3$ | Symmetric $CH_3$ stretching | 2872 | (*8, 9*) |
| $\nu_{as} CH_3$ | Asymmetric $CH_3$ stretching | 2960 | (*8, 9*) |
| $\delta CH_3$ | $CH_3$ bending | 1380 | (*8*) |
| $\nu_s CD_2$ | Symmetric $CD_2$ stretching | 2090 | (*10*) |
| $\nu_{as} CD_2$ | Asymmetric $CD_2$ stretching | 2195 | (*11*) |
| $\nu_s CD_3$ | Symmetric $CD_3$ stretching | ~2070-2080 | (*12*) |
| $\nu_{as} CD_3$ | Asymmetric $CD_3$ stretching | ~2210-2225 | (*12*) |
| $\nu OH$ | Alcohol/Phenol Hydroxyl stretching | 3110-3532 | (*13*) |

# Supplementary Figures

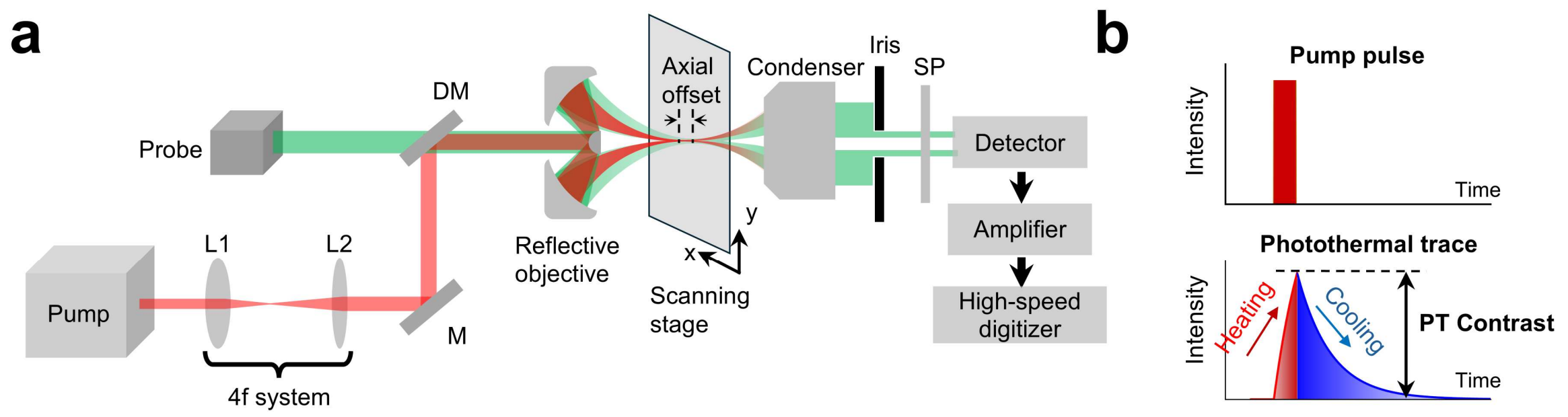

**Supplementary Figure 1. Details of MWIP microscope**. **(a)** Experimental set-up of MWIP microscope. A tunable mid-IR pump beam and a continuous-wave probe beam are combined by a dichroic mirror and focused onto the sample by a reflective objective. Axial offset is introduced using a 4f relay system. The transmitted probe beam is collected by a condenser, spatially filtered by an iris, spectrally filtered, and detected by a photodiode, followed by electronic amplification and high-speed digitization. DM, dichroic mirror; SP, short-pass filter. **(b)** Single-pulsed digitized photothermal signal and generation of photothermal contrast.

---

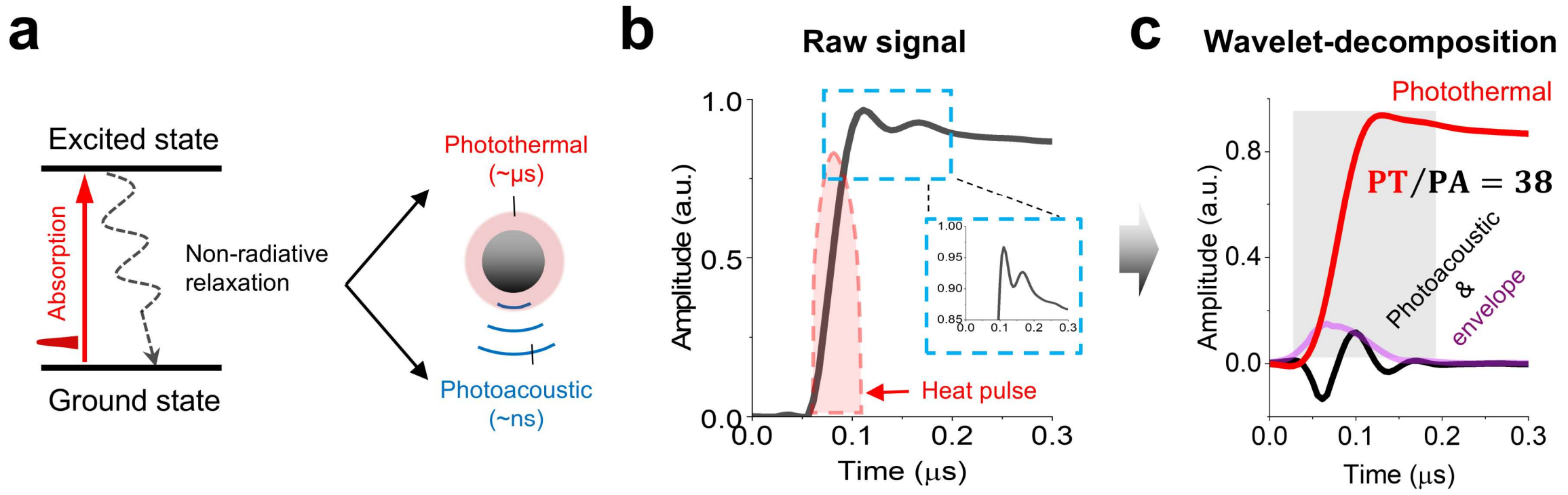

**Supplementary Figure 2. Photothermal signals dominate energy deposition compared to photoacoustic transients in nonradiative relaxation process.**
**(a)** Schematic of nonradiative relaxation leading to photothermal (μs) and photoacoustic (ns) pathways. **(b)** Raw transient signal showing overlapped photothermal and photoacoustic components. **(c)** Wavelet decomposition separating photothermal and photoacoustic contributions, yielding a photothermal-to-photoacoustic ratio of 38.

---

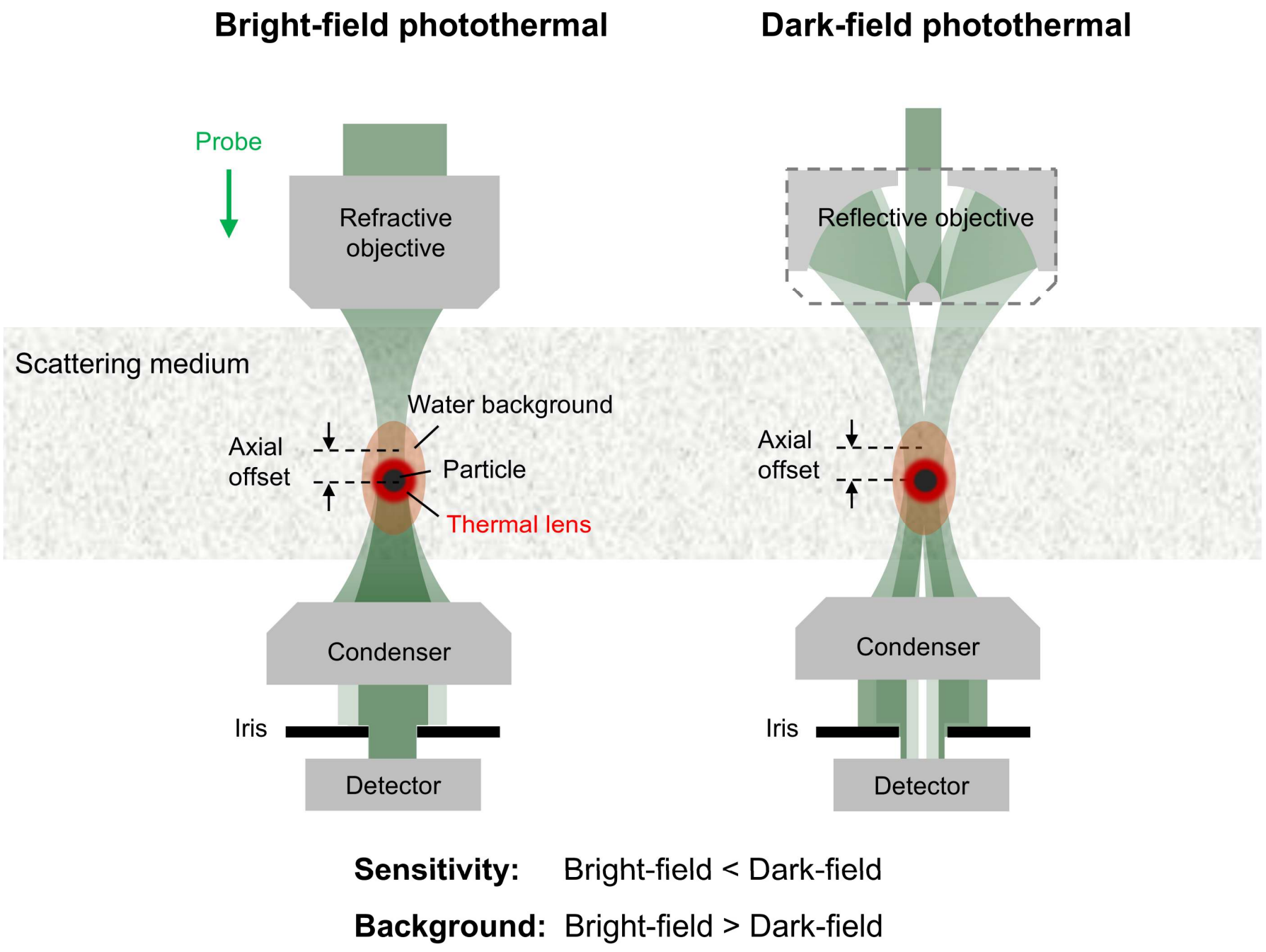

**Supplementary Figure 3. Schematic comparison of bright-field and dark-field photothermal detection.**

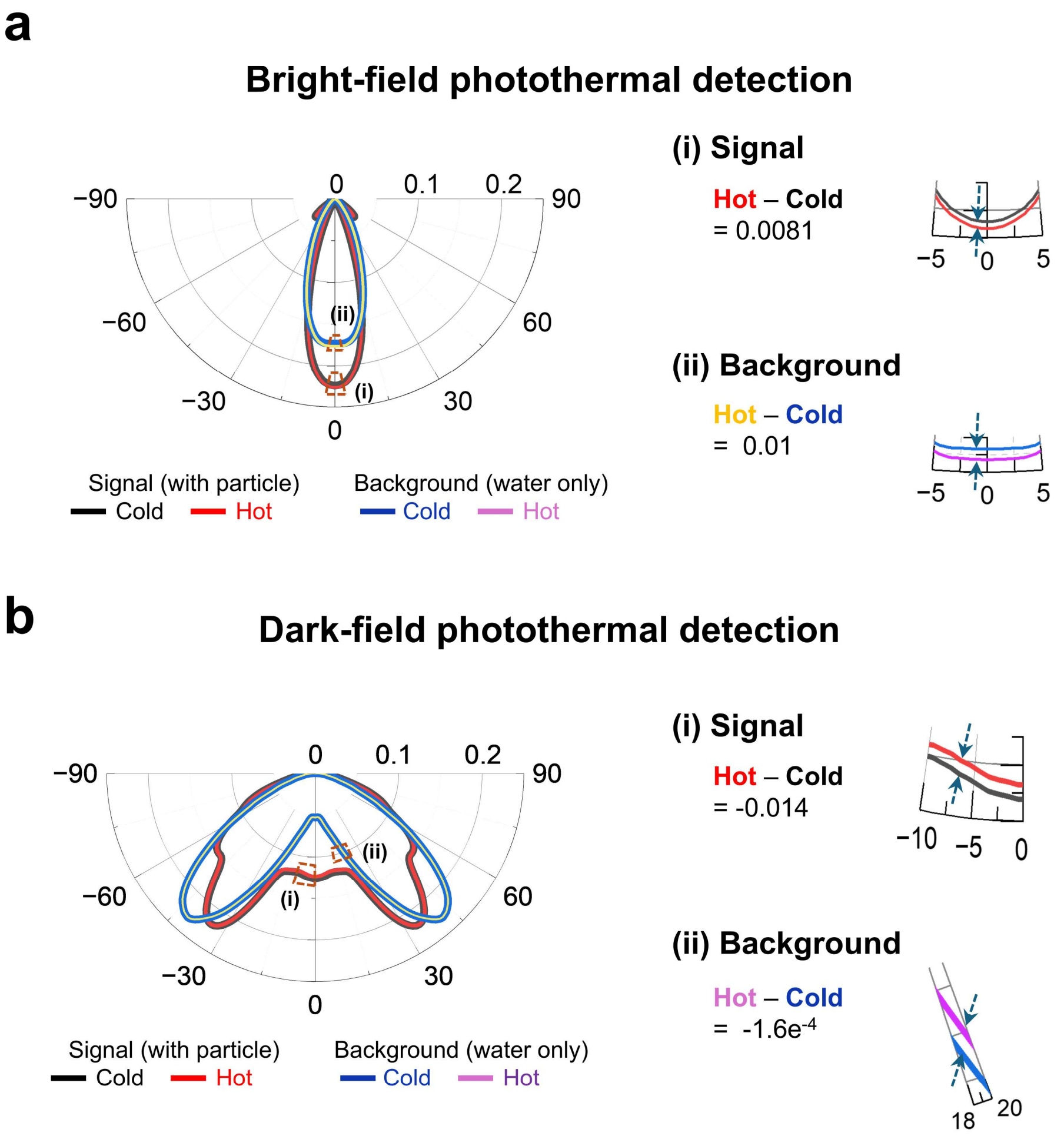

**Supplementary Figure 4. Far-field angular scattering patterns for photothermal detection. (a)** Bright-field and **(b)** dark-field photothermal detection schemes. Polar plots show the COMSOL simulated far-field angular scattering intensity for hot and cold states, including contributions from the particle signal and the water-only background. Right panels show the corresponding differential signals obtained by subtracting the cold state from the hot state (Hot − Cold). Insets provide magnified views of the regions marked **(i)-(ii)** in **(a)** and **(i)-(ii)** in **(b)**, highlighting the relative magnitude of the photothermal signal and background. Bright-field detection exhibits comparable signal and background levels, whereas dark-field detection strongly suppresses the background while preserving a pronounced photothermal signal. Thermal expansion of the particle reduces its scattering cross section, leading to a negative modulation depth in the dark-field configuration due to fewer scattered photons reaching the dark-field collection angles. In contrast, the bright-field photothermal scheme exhibits a positive modulation depth because more photons remain confined to the original forward-scattering directions.

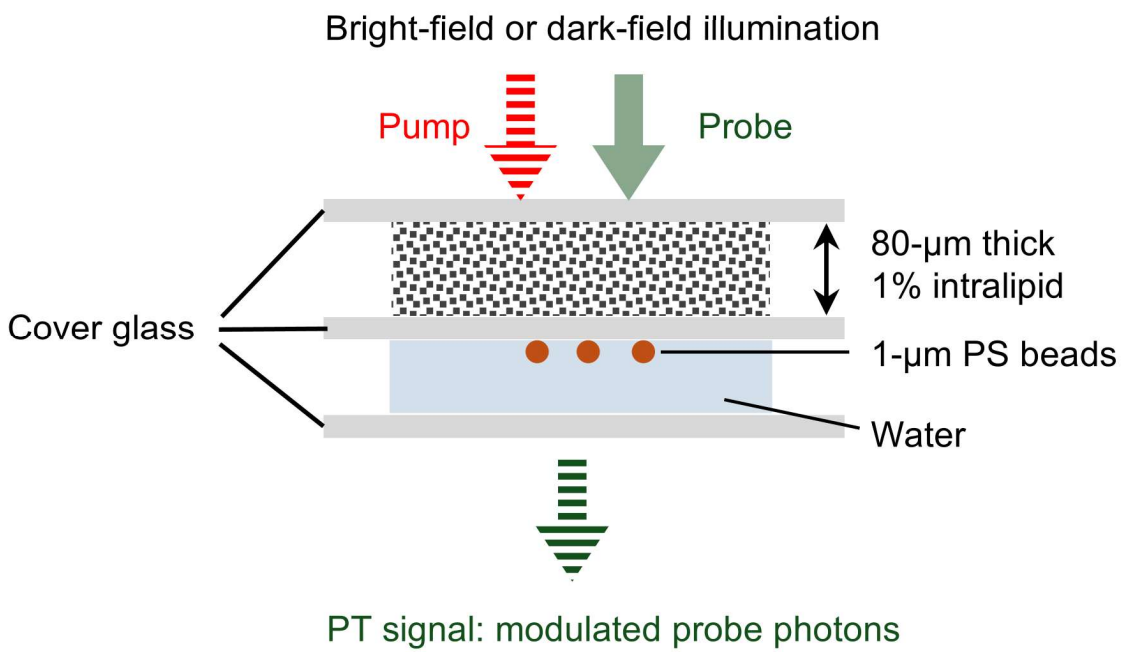

**Supplementary Figure 5. Bright-field (BF) and dark-field (DF) photothermal imaging schemes for 1-μm polystyrene (PS) particles in water under a scattering phantom (1% intralipid).** To enable a fair comparison between the two configurations, 1725-nm pump pulses were used, as this wavelength allows the use of both refractive and reflective objectives with the same focusing numerical aperture (NA = 0.5). Other experimental details are described in the **Methods**.

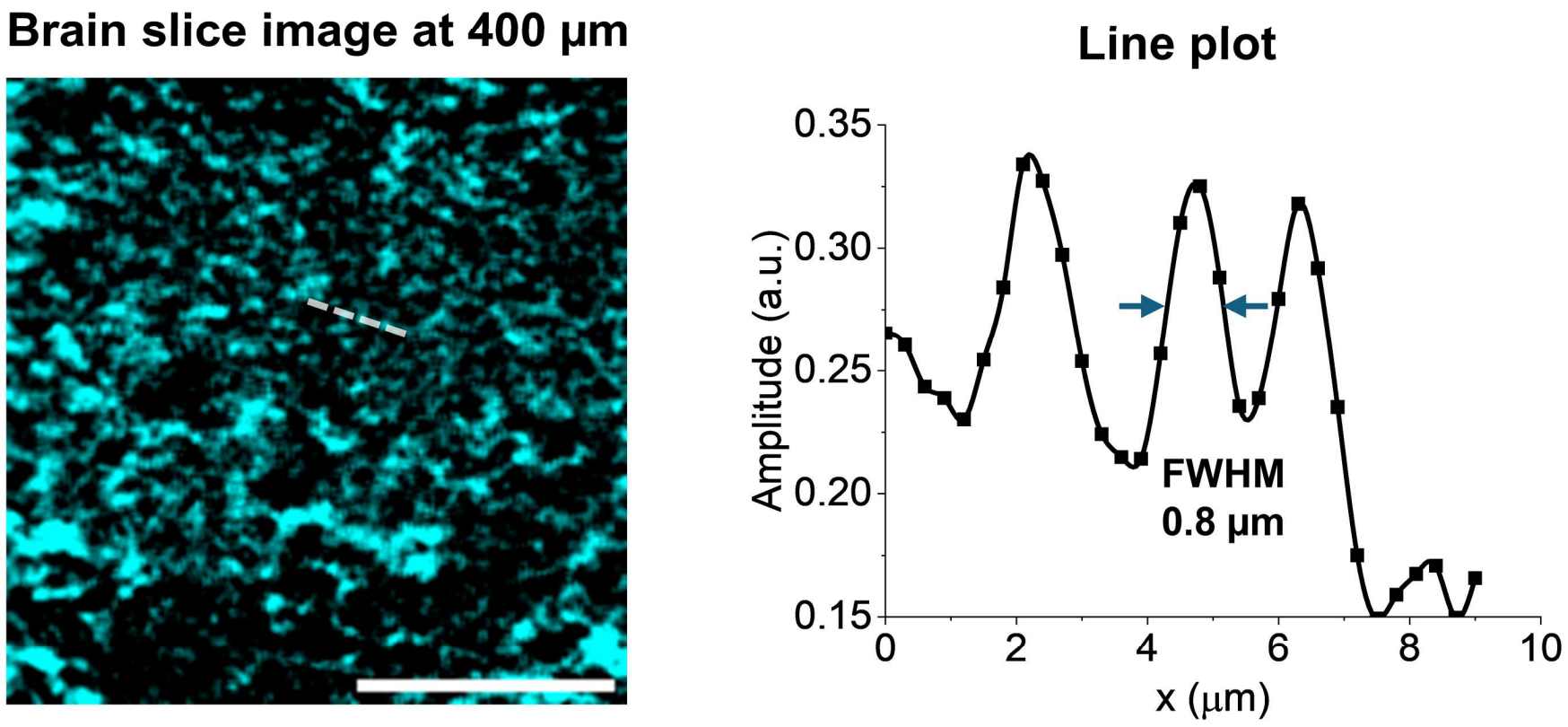

**Supplementary Figure 6. High-resolution MWIP imaging of a brain slice, resolving sub-micron chemical features at a depth of 400 μm using 2310 nm excitation**. Scale bar: 25 μm.

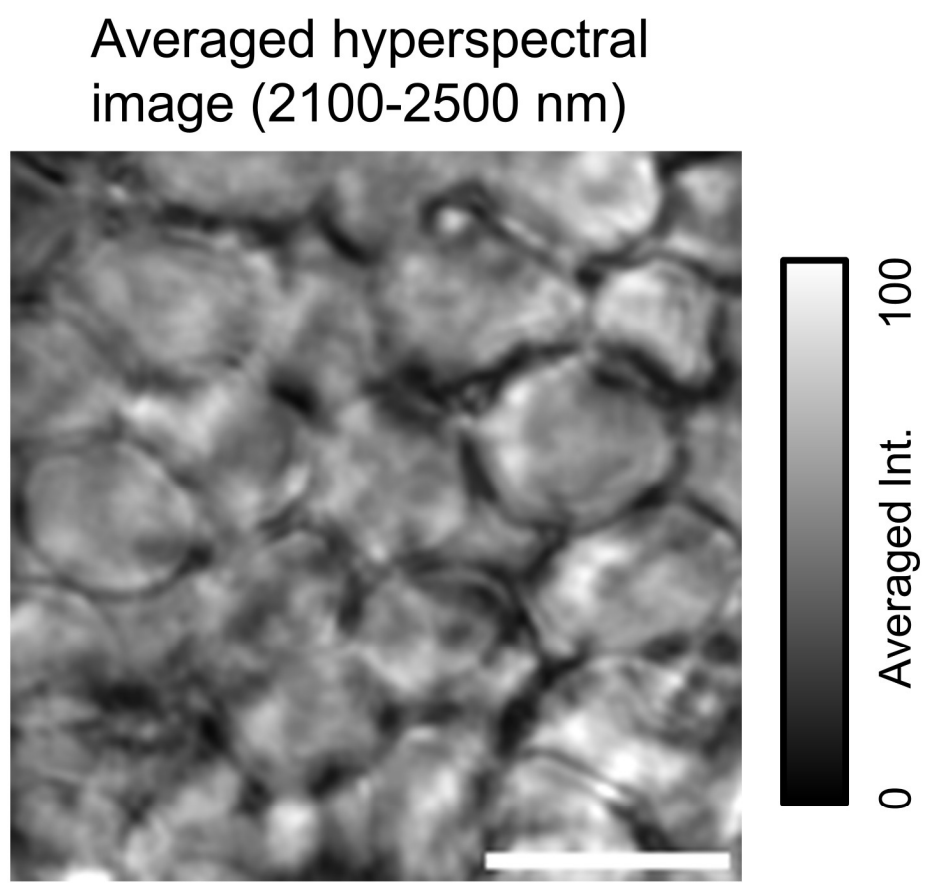

**Supplementary Figure 7. Averaged hyperspectral image (2100-2500 nm) of Fig. 4d to show the morphology of lipid-filled cartilage and drug content at a depth of 180 μm beneath the surface.** Scale bar: 30 μm.

---

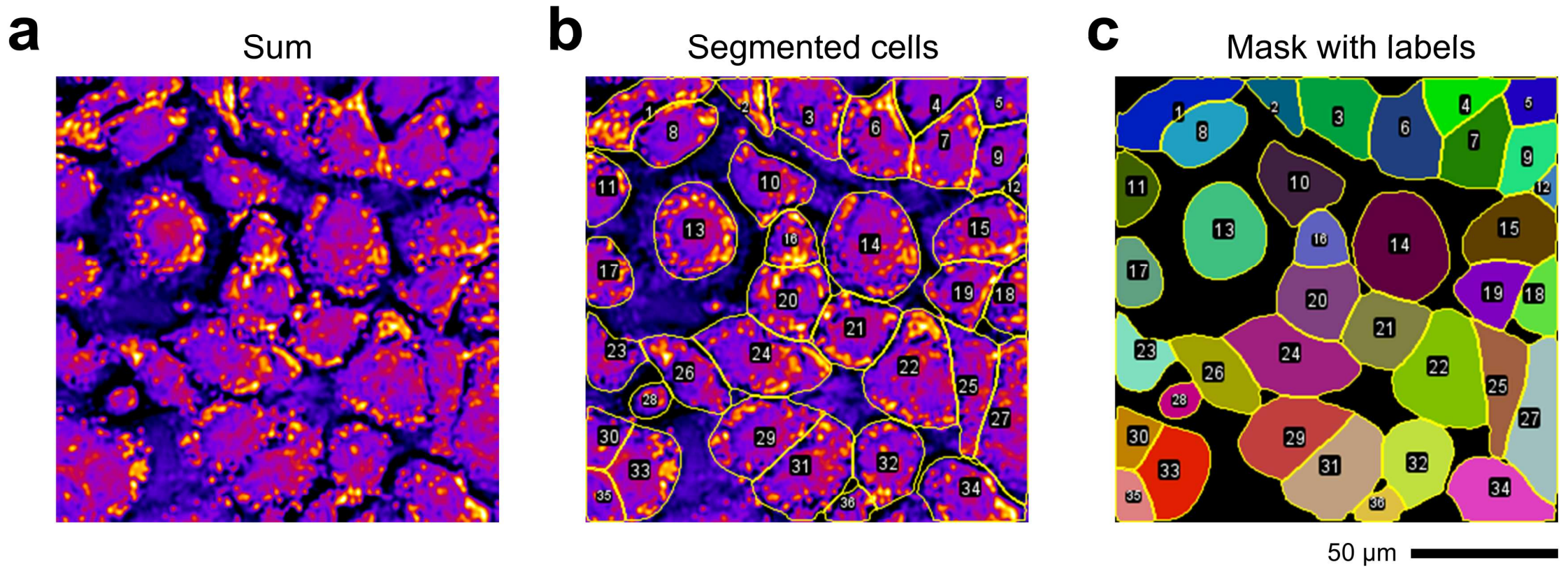

**Supplementary Figure 8. Deep-learning-based automated Cellpose cell segmentation for fatty acid uptake analysis.**
**(a)** Average hyperspectral image of 2D cultured cells. **(b)** Cellpose-generated cell boundaries. **(c)** Segmentation mask with labeled cell indices for downstream CD/CH quantification.

---